%% file: main.tex
\documentclass{article}
\usepackage{arxiv}
\usepackage[utf8]{inputenc} 
\usepackage[T1]{fontenc}    
\usepackage{hyperref}       
\usepackage{xurl}            
\usepackage{booktabs}       
\usepackage{amsfonts}       
\usepackage{nicefrac}       
\usepackage{microtype}      
\usepackage{lipsum}		    
\usepackage{graphicx}
\usepackage{natbib}
\usepackage{doi}
\usepackage[super]{nth}
\usepackage{threeparttable}
\usepackage{multirow}
\usepackage[table]{xcolor}

\title{Filtering Eye-Tracking Data From an EyeLink 1000: Comparing Heuristic, Savitzky-Golay, IIR and FIR Digital Filters}

\author{ {\hspace{1mm}Mehedi H. Raju}\thanks{corresponding author} \\
	Texas State University\\
    601 University Drive\\
    San Marcos, Texas, 78640, USA\\
	\texttt{m.raju@txstate.edu} \\
	\And
	{\hspace{1mm}Lee Friedman} \\
	Texas State University\\
    601 University Drive\\
    San Marcos, Texas, 78640, USA\\
	\texttt{lfriedman10@gmail.com} \\
	\And
	{\hspace{1mm}Troy M. Bouman} \\
	Department of Mechanical Engineering-\\Engineering Mechanics\\
    Michigan Technological University\\
    1400 Townsend Dr.\\
    Houghton, MI 49931, USA \\
	\texttt{tmbouman@mtu.edu} \\
	\And
	{\hspace{1mm}Oleg V. Komogortsev} \\
    Texas State University\\
    601 University Drive\\
    San Marcos, Texas, 78640, USA\\
	\texttt{ok11@txstate.edu} \\
}

\renewcommand{\shorttitle}{Comparing Heuristic, Savitzky-Golay, IIR and FIR Digital Filters}

\begin{document}
\maketitle

\begin{abstract}

In a previous report \citep{raju_signal} we concluded that, if the goal was to preserve events such as saccades, microsaccades, and smooth pursuit in eye-tracking recordings, data with sine wave frequencies less than 100 Hz (-3db) were the signal and data above 100 Hz were noise.  We compare 5 filters in their ability to preserve signal and remove noise. Specifically, we compared the proprietary STD and EXTRA heuristic filters provided by our EyeLink 1000 (SR-Research, Ottawa, Canada), a Savitzky-Golay (SG) filter, an infinite impulse response (IIR) filter (low-pass Butterworth), and a finite impulse filter (FIR). For each of the non-heuristic filters, we systematically searched for optimal parameters.  Both the IIR and the FIR filters were zero-phase filters.  Mean frequency response profiles and amplitude spectra for all 5 filters are provided.  In addition, we examined the effect of our filters on a noisy recording.  Our FIR filter had the sharpest roll-off of any filter. Therefore, it maintained the signal and removed noise more effectively than any other filter. On this basis, we recommend the use of our FIR filter.  Several reports have shown that filtering increased the temporal autocorrelation of a signal. To address this, the present filters were also evaluated in terms of autocorrelation (specifically the first 3 lags).  Of all our filters, the STD filter introduced the least amount of autocorrelation.

\end{abstract}


\input{sections/01-intro.tex}
\input{sections/02-method.tex}
\input{sections/03-result.tex}
\input{sections/04-discussion.tex}

\section*{Acknowledgments}
This work was funded by grant from the NSF (1714623) (PI: Oleg Komogortsev).


\section*{Conflict of interest}
The authors declare no conflict of interest. The funders had no role in the design of the study; in the collection, analyses, or interpretation of data; in the writing of the manuscript, or in the decision to publish the results.

\bibliographystyle{unsrtnat}
\bibliography{Fourier.bib} 

\end{document}

%% file: sections/01-intro.tex
\section{Introduction}
\label{intro}

According to \cite{raju_signal}, for the study of saccades, microsaccades, and smooth pursuit, frequency components above 100~Hz can be considered as noise.  This was based on several forms of analysis: (1) a visual analysis of different frequency components, (2) an analysis of the percent of variance accounted for by various frequency bands, and (3) a detailed study of the effect of low-pass filtering on saccade peak-velocity.  Based on the results of these analyses we concluded that signals comprised of sine-wave frequencies below 100 Hz are essential for visualizing eye-movement events and measuring the peak velocity of saccades.  Sine-wave frequencies above 100 Hz can be considered noise (see also \cite{bahill80Hz, mack}).

In 1993, \cite{stampe} suggested two ``heuristic'' filters that were designed for the video-oculography system.  One was labeled standard (STD) and was labeled extra (EXTRA).  Several manufacturers\footnote{SR-Research (Ottawa, Canada), Tobii (Stockholm, Sweden), and the XVIEW system from SMI.} have, over the years, employed these filters. At some (unknown) point in time, SR-Research modified both of the original heuristic filters. However, the date of the change and the nature of the modifications are proprietary.

\cite{mack} evaluated moving average (MA) \citep{movingavgM}, Savitzky-Golay (SG) \citep{savitzkyGolayM} and low-pass Butterworth (BW)\citep{butterworthM} filters. Both the MA filter and the SG filter are FIR-style filters \citep{movingavg, sg, mack}. They compared the performance of both FIR (MA and SG) and IIR (Butterworth) filters on saccadic movements\footnote{It is important to note that \cite{mack} tested all of their filters on synthetic saccades.}. These authors suggested that for 1000 Hz data, the BW performed better than the various MA or SG filters examined. 
Based on their analysis, we decided to further study SG,  and Butterworth filters (IIR-type). In addition, we also evaluated a standard FIR low-pass filter not evaluated by \cite{mack}\footnote{
\cite{das1996measuring} evaluated the effectiveness of combined median and moving-average filters to reduce velocity noise in smooth pursuit vestibular eye movements.}.

The main objective of this study is to determine the most effective filtering approach for preserving eye-movement signals below 100 Hz and eliminating frequency components above 100 Hz that are considered noise.  We compare the effectiveness of heuristic and digital filters in terms of their ability to preserve signals and eliminate noise. The key analysis is the comparison of the frequency response curves for all filters.  In addition, since it is well established that filtering typically increases temporal autocorrelation \citep{autocorrelation}\footnote{See also Friedman et al, at \url{https://digital.library.txstate.edu/handle/10877/15303}, ActualArXivSubmission.pdf}, we also compare the autocorrelation functions for signals processed with all filter types.

%% file: sections/02-method.tex
\section{Methods}
\label{method}

\subsection{Subjects}
\label{data}
A total of 23 subjects were recruited (M-17, F-6), Median age was 28 (range: 20 to 69 yrs).  A majority (N=14) of participants had normal vision, while 9 subjects needed vision correction. The participants were recruited from laboratory personnel, undergraduate students taking a computer programming course, and friends of the experimenters. The study was approved by the Texas State University institutional review board and all participants provided informed consent.

We report on two datasets, the first dataset is labeled as the ``Fixation'' dataset. This dataset originally had data from 15 subjects. However, due to blinks and other technical issues, only data from 9 subjects were  analyzed. The second dataset (“RS”), contained data when subjects viewed a random saccade task. The RS dataset consisted of 9 subjects.

\subsection{Data Collection}

During the data collection process, the participants were positioned at a fixed distance of 550 millimeters from a 19'' (48.26 cm) computer screen (474$\times$297 millimeters, resolution 1680$\times$1050 pixels), where they were presented with visual stimuli. The data was captured using a tower-mounted EyeLink 1000 eye tracker (SR Research in Ottawa, Ontario, Canada) and operated in a monocular mode to record the movement of the dominant eye. The participants' dominant eye was identified using the Miles method \citep{MilesTest}. The sampling rate was 1000 Hz. For each subject, there were three fixations recorded: (1) Unfiltered; (2) STD filtered, and (3) EXTRA filtered.

For the fixation task, participants were presented with a white circle with a diameter of $0.93^o$ as a visual stimulus. The circle was positioned at a distance of 3.5$^o$ above the primary position, at the horizontal middle of the screen. Participants were instructed to maintain their gaze on the stationary point for 30 seconds.

For the random saccade task, the participants were instructed to follow the same target that moved randomly across the display monitor, ranging from ± 15$^o$ and ± 9$^o$ of visual angle in the horizontal and vertical directions respectively. The minimum amplitude between adjacent target displacements was 2$^o$ of visual angle. The target positions were randomized for each recording to ensure uniform coverage across the display. The delay between target jumps varied between 1 second and 1.5 seconds, chosen randomly from a uniform distribution. The random saccade task lasted for 30 seconds\footnote{
For more details about subjects and data collection procedures see \cite{raju_signal}.}.

\subsection{Signal Processing of Fixation Data} 
\label{sigpro}
All fixation recordings lasted 30 seconds (30,000 samples).  We used these fixation periods to create amplitude spectra and to determine the frequency response of several filters discussed below.  The segment selection was a two-step process. In the first step, we calculated the velocity with a six-point difference approach using $ velocity = (x_{t+3} - x_{t-3})/dt$ \citep{bahilltwopoint}. We then screened the recordings of each subject to find the maximum number of segments of length 2048 samples that did not have any velocity above 25 deg/sec.  We rejected any segments which contained velocities above 25 deg/sec to reduce the possibility of saccades or other fast events in our segments. For four of the 16 subjects, we could not find a single segment of 2048 samples that met our criteria. For the remaining subjects, we found 1 to 4 segments.  We  use these 2048-sample segments for our Fourier analysis. Using the fast Fourier transform (FFT), the ratio of the sampling rate to the segment size (i.e., block size) determines the frequency resolution.  For 2048-sample segments, we could discriminate 1024 different frequencies from 0 to 500 Hz.  Since these analyses were quite noisy, we decided to break down the 2048-sample segments into 8 256-sample segments.  This would still give us reasonable frequency precision of approximately 4 Hz, and it would produce more segments to average (we had 27 2048-sample segments and these produced 216 segments of 256 samples across which to average).  Note that we did our averaging using the magnitude  spectra rather than averaging the complex FFT data.

\subsection{Digital Filter Design}
\label{filterdesign}

\begin{table*}[htbp]
\centering
    \caption{Digital filter specification}
    \begin{tabular}{ll}
    \hline
Filter name                      & Filter characteristics                       \\ \hline
Savitzky-Golay (SG)              & Window length=11, polynomial order=2         \\ \hline 
Infinite impulse response (IIR)  & Butterworth type, Order=7 \\
 & Cut-off (-3dB)  = 100, Zero-phase \\ \hline
Finite impulse response (FIR)    & Number of coefficients, taps = 80\\ 
 & Cut-off (-3dB) = 100, Zero-phase\\ \hline
\end{tabular}
\label{tab-filters}
\end{table*}

As we want to retain the frequency components below 100~Hz and remove noise above this, we chose a cut-off of 100~Hz. When we refer to a ``cutoff'' frequency, we are referring to the -3dB (dB = decibels) point, which is standard in the signal processing literature.  At the -3dB point, signals are reduced by 50\%.  

Before choosing a final set of parameters (order and window length) for the SG filter, we examined the frequency response of  SG filters with orders from 2 to 9 and window lengths from 5 to 91 (odd numbers only).  For implementing the SG filter we used the python built-in function from Scipy\citep{scipy}.
We were looking for filters that had a -3dB point near 100 Hz.  We found 4 sets of parameters that met this criterion.  All of these 4 parameter sets produced similar frequency responses so we chose the parameter set with the lowest order and window size (order 2, window size 11).  It is generally known that SG filters have substantial ringing in the stop band \citep{sg-fr} (see Fig. \ref{fig:allfreqz}).  As we tested various potential parameter settings we noted that higher orders produce fewer and wider ringing lobes, Increasing the window size produces more, smaller (in terms of dB), and narrower ringing lobes.

For our IIR filter, we chose a Butterworth low-pass filter with order = 7 and a cutoff of 100 Hz. We chose order 7  because it has a steep roll-off and appeared to be stable.  We formally checked for the stability of the IIR filter with the unit-circle test\footnote{For this test we used MATLAB function \texttt{isstable}.}.

FIR filters are always stable. For our FIR filter, we chose 80 taps. We determined this number of taps ($N_{taps}$) based on the following formula from \citep{digsig}.

$N_{taps} \approx \frac{2}{3}\cdot  log_{10}\cdot  [\frac{1}{10(\delta_{1} \delta_{2} )}]\cdot  \frac{fs}{\Delta f}$

Here, $N_{taps}$ = number of taps (filter order) \newline
$\delta_{1}$ = the ripple in passband \newline
$\delta_{2}$ = the suppression in the stop band \newline
$fs$ =the sampling rate \newline
$\Delta f$ = the transition width.

For the IIR filter, we used Butterworth (BW) low-pass filter.  The BW filter is maximally flat in the pass band and has no ripples in the stop band. Also, BW filters do not have any linear phase response in contrast to finite impulse response filters \citep{mack}. 
Mack et al go on to say:
\begin{quote}
    ``A general observation from the best filter list is the increasing prevalence of BW filters at higher sampling rates, ending in a total absence of other filter types at 1 kHz. This can be explained by considering the smoothness of the signal. At higher sampling rates more noise is present in the data. Such high-frequency noise can be more efficiently suppressed by the steeper roll-off of the BW filters compared to the two FIR filters, resulting in a smoother signal.\ldots''\citep{mack}, page 2159.\end{quote}

Typically, low-pass digital filters can have phase and delay effects.  Both our FIR and IIR filters are zero-phase and zero-delay.  A zero-phase filter can be constructed by first passing the signal through the filter in the forward direction, then reversing the filtered sequence and running it back through the filter. This process doubles the order of the filter and removes both phase and delay effects.  

Table \ref{tab-filters} represents a list of filters we employed along with their characteristics.

\subsection{Estimation of Filter Frequency Response}
\label{freqResponse}
We wanted to estimate the frequency response of each of our five filters.  For the designed FIR and IIR filters, the frequency response could be computed directly from the filter coefficients. Due to the zero-phase design, we directly obtained the frequency response after doubling the filter order.

For all filters, we start with an FFT of unfiltered data.  Let us label this FFT as ``A''. We also calculated the FFT for each filtered dataset. Let us label the FFT of the filtered signal as ``B''. We compute the filter frequency response C by ratio method, $C~=~\frac{B}{A}$.

\subsection{Fourier Analysis of Fixation before and after filtering}
\label{fourieranlysis}

To further study the effects of the filters, we filtered our fixation data with all five filters.  Then we computed the amplitude spectrum of all of these filtered (and unfiltered) signals using FFT\citep{FFT}.  These amplitude spectra show how the filters affected the amplitude of the signal that remains after filtering.

The input fixation data consisted of 216 blocks, each 256 samples long, as mentioned earlier. In the first step, each segment was detrended with a \nth{2} order polynomial. The residuals of these polynomials have a mean of zero. A Hanning window is then applied to each fixation segment. We then perform an FFT of each detrended and windowed fixation segment  The resulting spectra have a frequency resolution of $\approx$ 4 (3.90625) Hz. With a sample rate of 1000 Hz, spectra can only be calculated from 0 to 500 Hz \citep{shanon}.  With  a 3.90625 Hz resolution, we end up with spectra that are 128 points long. These amplitude spectra were averaged across all 216 fixation segments.  This produced a relatively clean amplitude spectrum.

\subsection{Study of the Effects of Filtering on Temporal Auto-correlation}
\label{ACF}
It is known that low-pass filtering can increase temporal autocorrelation \citep{autocorrelation}(also see footnote 4).  We thought it would be useful to examine the effects of our filtering scheme on temporal autocorrelation.  For each fixation segment, we calculated the autocorrelation function (ACF) out to 5 lags.  For each filtered set of autocorrelations, we plotted the median.  Since Pearson r correlation coefficients are not on a linear scale, prior to statistical testing all of the ACF estimates were transformed using a Fisher-Z transformation.  For the first 3 lags, we tested the statistical significance of the differences between autocorrelation estimates for unfiltered and filtered data.  We used the Friedman test.  We followed up a statistically significant Friedman test with a multiple comparisons comparing all filter levels.   Controlled for multiple comparisons employed the Tukey HSD test.

\subsection{Study of the Effects of Filtering on Positional Signal and Velocity}
\label{method-exemplar}

To illustrate the effect of our filters on actual position traces, we studied the effect of three filtes (SG, IIR, FIR) on an unfiltered exemplar segment with a saccade. Out of 9 subjects from the ``RS'' dataset, we chose the subject with the noisiest recording.  We chose to illustrate instantaneous velocity ($velocity =\frac{x_{t}-x_{t-1}}{dt}$) for this case because it is the noisiest velocity calculation. 

%% file: sections/03-result.tex
\section{Results}
\label{result}

\subsection{Analysis of Filter Frequency Response}\label{fr}

\begin{figure*}[htbp]
\centering
\includegraphics[width=0.9\textwidth]{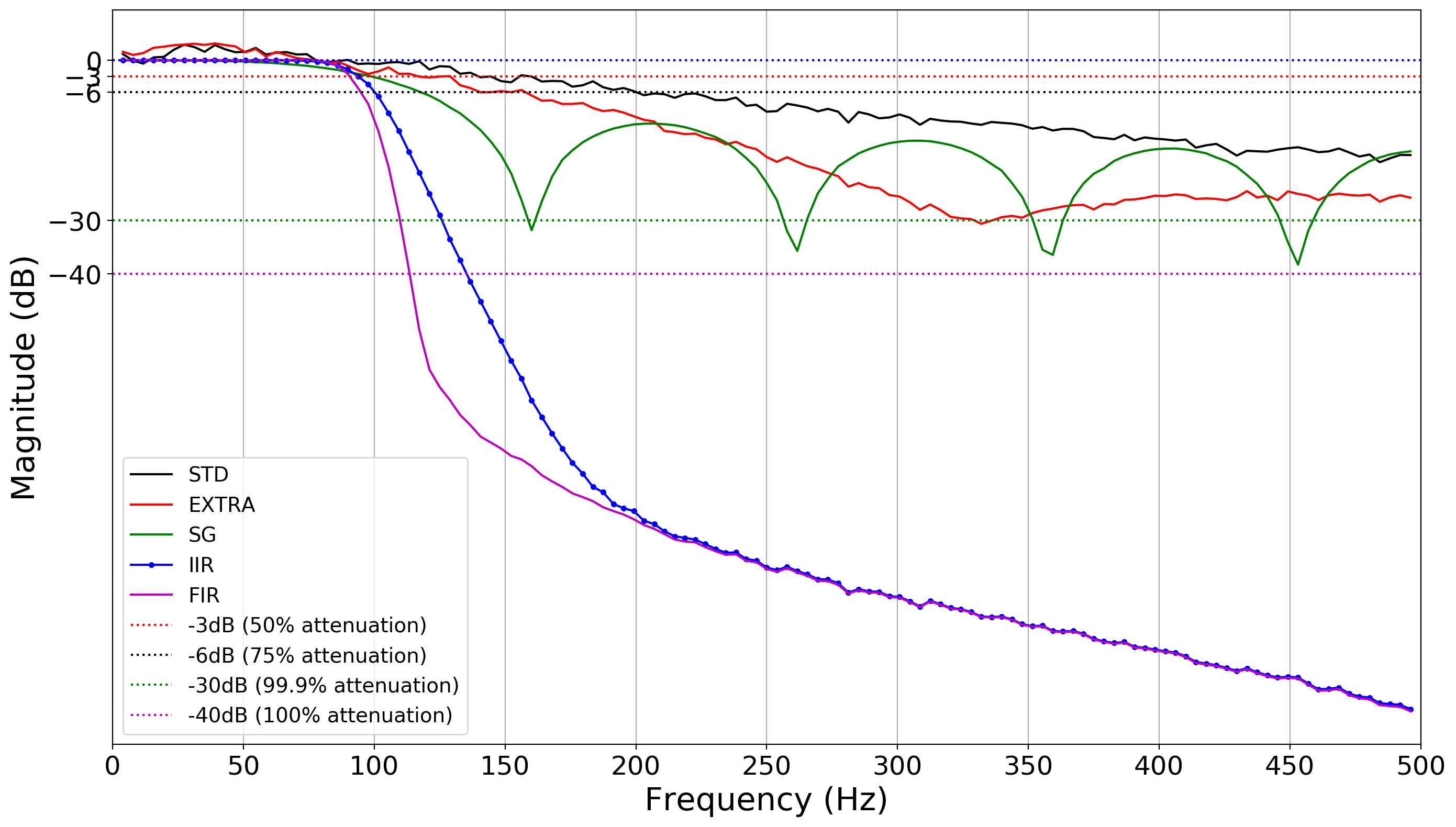}
\caption{Frequency response of all the filters (EyeLink heuristic filters and Digital filters). At -3dB signals are reduced by 50\%, at -6dB signals are reduced by 75\%, and so on as mentioned in the legend.}
\label{fig:allfreqz}
\end{figure*}

In Fig. \ref{fig:allfreqz}, we present the frequency response of STD, EXTRA, SG, IIR, and FIR filters. The Y-axis of the plot is in decibels (dB, reference = 1). The black line represents the frequency response of the STD filter. The red line represents the frequency response of the  EXTRA filter.  The green line represents the frequency response of the SG filter. The blue and magenta lines represent the frequency response of the IIR and FIR filters, respectively.  It appears that all of the filters do a good job of preserving the signal (frequencies less than 100 Hz).  However, the filters differ substantially in the degree to which they remove noise-related frequencies.  It is obvious that the FIR and IIR filters have much sharper roll-offs than the other filters.   Only the IIR and FIR filters ever achieve -30 db (0.1 \% of signal amplitude remaining).  IIR reaches this point at 127 Hz and FIR at 110 Hz.  The SG filter reaches this point at 158 Hz, and the EXTRA filter reaches this point at 332 Hz.  The STD never reaches this level of reduction.

Obviously, the digital filters do a much better job of reducing higher frequencies without affecting the lower frequencies.  The heuristic filters remove high-frequency signals much more slowly than the digital filters. The maximum amplitude of noise remaining at 500 Hz is -18 dB (1.58 percent of the signal remaining) for the STD filter whereas, for the EXTRA filter, it is -25 dB (0.32 percent of the signal remaining).  The signal is effectively reduced to $0^o$ amplitude (-40 dB) at 135 Hz for the IIR filter and at 114 Hz for the FIR filter.  The frequency response of the FIR filter is steeper than all other filters.

\subsection{Fourier analysis of the unfiltered and filtered signals}
\label{Fourier}

\begin{figure*}[htbp]
\centering
\includegraphics[width=0.9\textwidth]{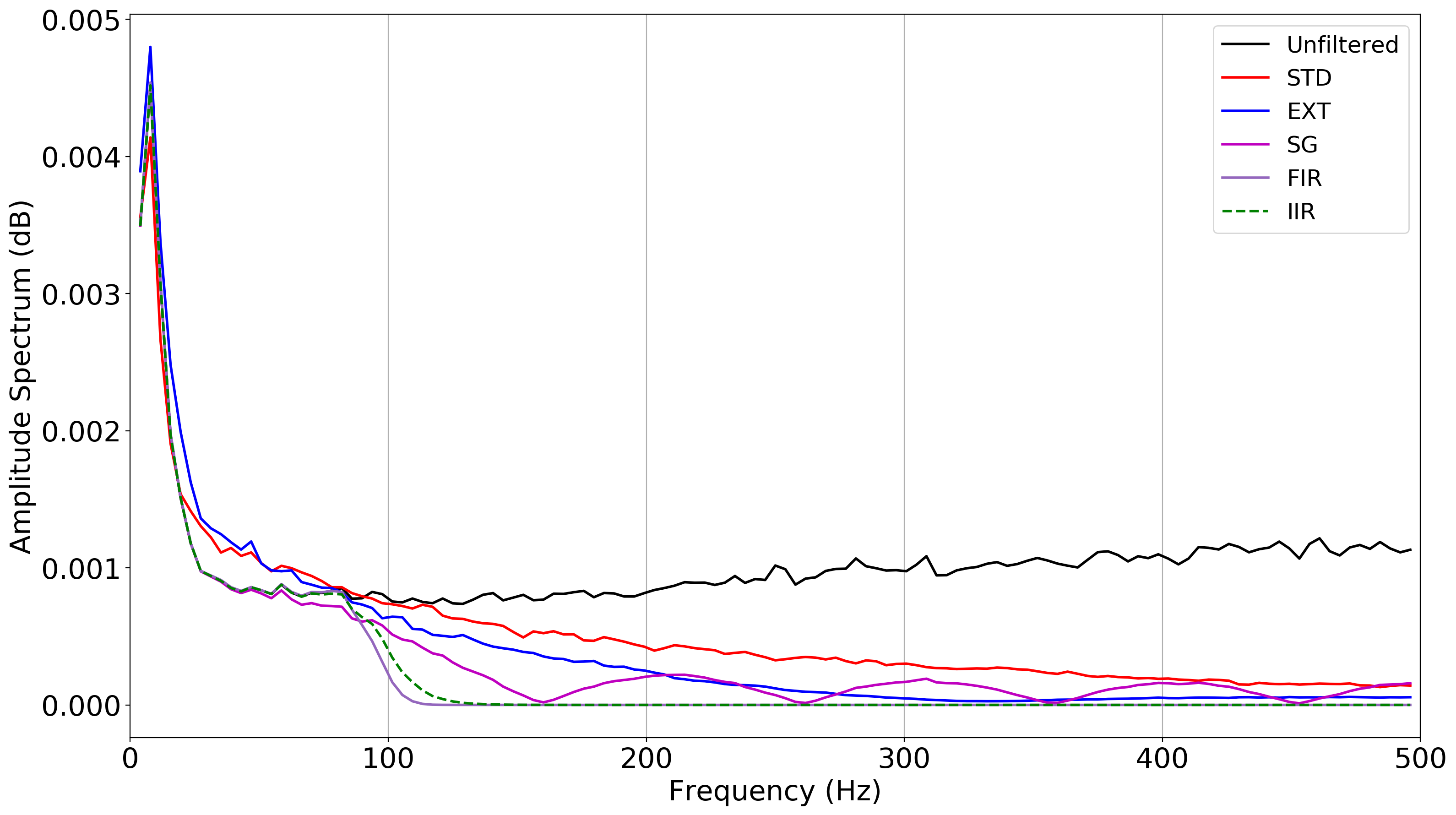}
\caption{Amplitude spectrum of the unfiltered signal and all of the filters evaluated in this report. Segments were chosen as described above.}
\label{fig:fft}
\end{figure*}

In Fig. \ref{fig:fft}, we present the average amplitude spectra for all 5 signal types.
Amplitudes for all signals are much higher in very low frequencies (1-30~Hz).  All filtered signals have less amplitude at above 100~Hz.  The amplitude of the unfiltered signal reaches a minimum of around 150~Hz and then the amplitude of noise frequencies actually increases as frequencies approach 500~Hz. The STD-filtered signal has a gradual decline in amplitude from about 50~Hz to 500~Hz.  The EXTRA-filtered signals remove substantially more noise frequencies than the STD filter. The SG filter has marked ringing in the stop band.  The amplitude of the IIR-filtered signal drops sharply at about 75 to 150~Hz and remains essentially 0.000 above 150~Hz.  The amplitude of the FIR-filtered signal drops sharply at about 75~Hz to 120~Hz and remains essentially 0.0 above 125~Hz.

\subsection{Effect of Filtering on Temporal Auto-correlation}
\begin{figure*}[htbp]
\centering
\includegraphics[width=0.7\textwidth]{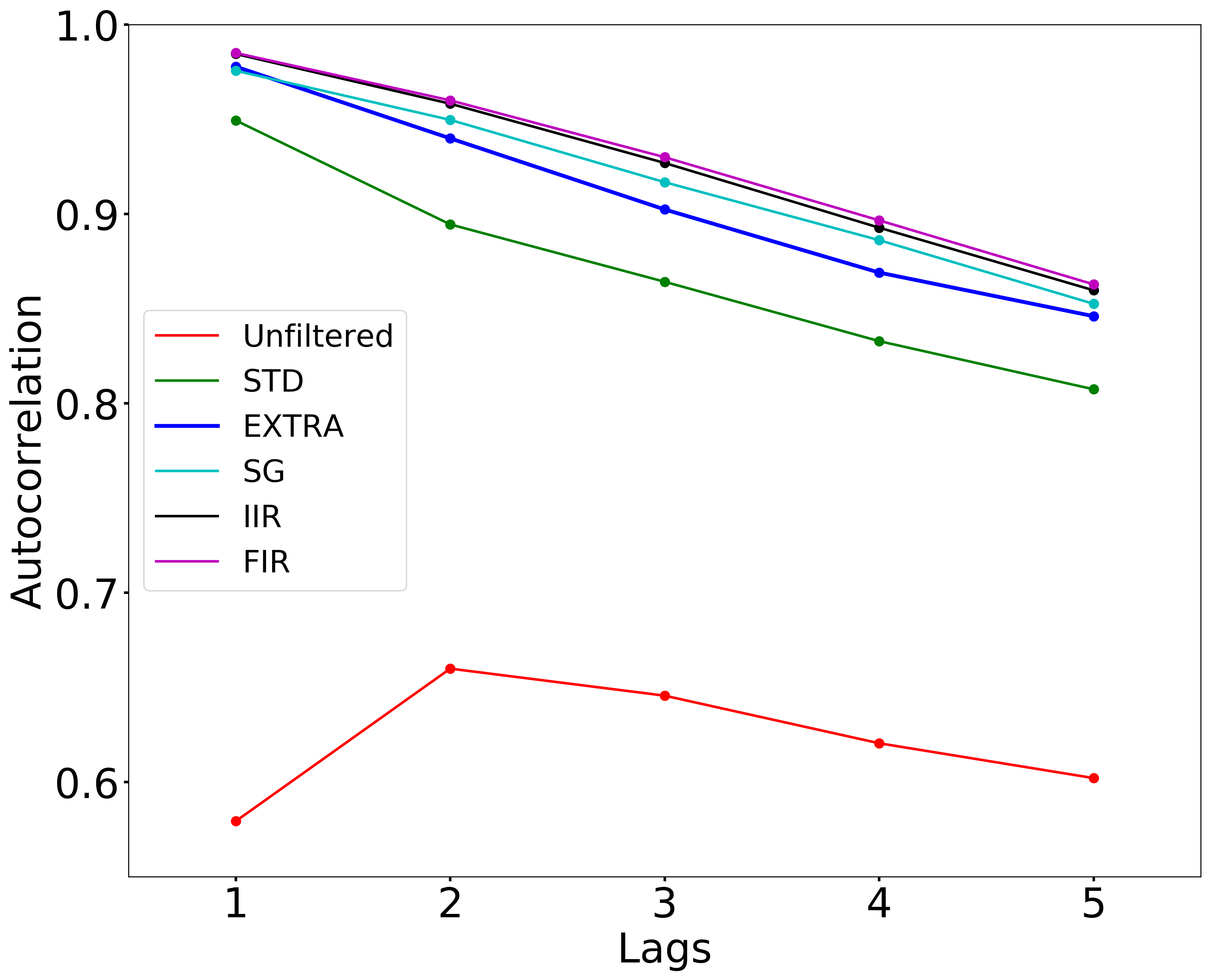}
\caption{Effect of filtering on median temporal autocorrelation for unfiltered and filtered fixation segments.}
\label{fig:acf}
\end{figure*}

The median ACF for the first 5 lags is plotted for unfiltered and filtered signals in Fig \ref{fig:acf}. In the unfiltered condition, the median lag 1 temporal autocorrelation was $\approx$ 0.579, and of a total of 216 segments, 178 were statistically significant at the $p<0.05$ level.
Although the unfiltered data reveals moderately strong temporal autocorrelation, it is clear that the filters do indeed introduce more temporal autocorrelation.  For all 216 fixations, filtered at all 5 levels, all 216 segments had a lag 1 temporal autocorrelation that was statistically significant at a $p~<~0.0001$ level. 

\begin{figure*}[htbp]
\centering
\includegraphics[width=0.95\textwidth]{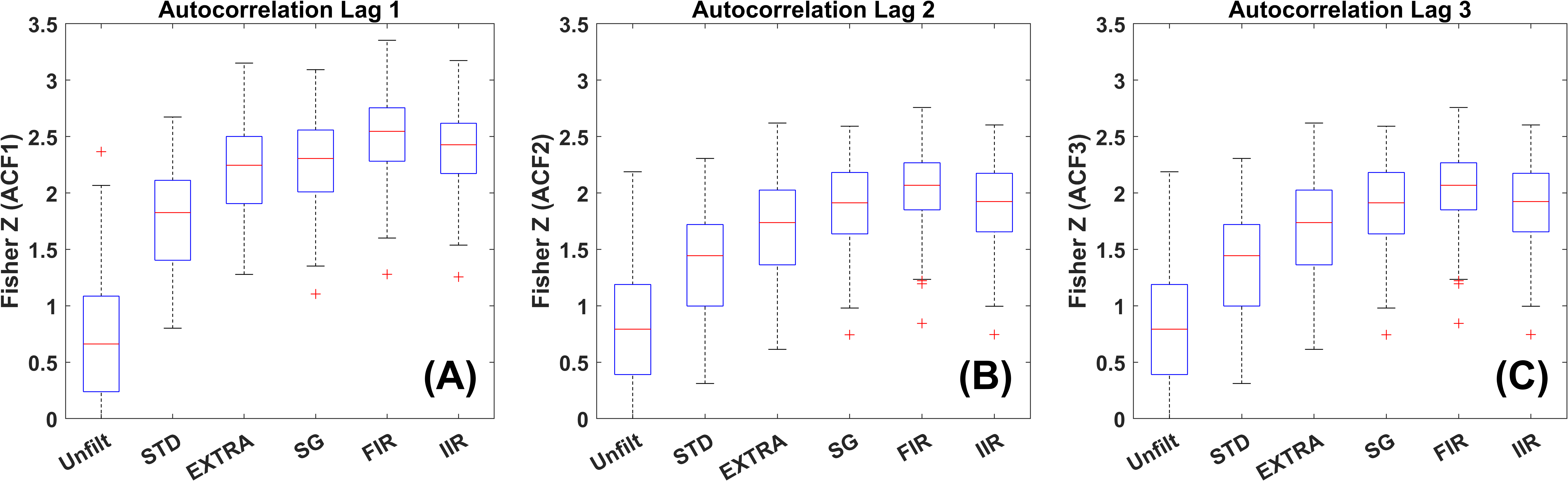}
\caption{Analysis of autocorrelation results. Values plotted are Fisher Z transformed values from the original autocorrelations.  Three box-plots that compare all filters.  (A) Box-plots represent ACF lag 1. (B) Box-plots represent ACF lag 2. (C) Box-plots represent ACF lag 3.}
\label{fig:boxplot}
\end{figure*}

In Fig.\ref{fig:boxplot}, we present boxplots for the Fisher-Z transformed values.  In (A), we present the results for lag 1, (B) for lag 2, and (C) for lag 3.  P-values from the Friedman tests were all statistically significant ($p < 0.0001$), (see Table 2).  The results of  all possible comparisons are presented in Table 2.  Comparisons that were not statistically significant are highlighted in yellow.  At a glance, it is clear that almost every comparison was statistically significant.  For lags 1 and 2, the EXTRA and SG were not statistically significantly different.  For lags 1 and 3, EXTRA and IIR were not statistically significantly different.  For lag 2 only the SG and the IIR were not statistically different.

\begin{table*}[htbp]
\label{tab:multi}
\centering
\begin{threeparttable}
\caption{Testing the Effects of filtering on Temporal Autocorrelation: Multiple Comparison Statistics$^*$}
\begin{tabular}{|l|l|ll|cc|cc|}
\hline
\multirow{3}{*}{\nth{1} Filter} & \multirow{3}{*}{\nth{2} Filter} & \multicolumn{2}{c|}{ACF 1} & \multicolumn{2}{c|}{ACF 2} & \multicolumn{2}{c|}{ACF 3} \\ \cline{3-8} 
~&  & \multicolumn{2}{c|}{$\tilde{\chi}^2 = 764.51^\dag$ } & \multicolumn{2}{c|}{$\tilde{\chi}^2 = 706.7^\dag$}  & \multicolumn{2}{c|}{$\tilde{\chi}^2 = 642.63^\dag$} \\ \cline{3-8} 
~  &  & \multicolumn{1}{c|}{Difference}  & P-value  & \multicolumn{1}{c|}{Difference} & P-value & \multicolumn{1}{c|}{Difference} & P-value \\ \hline
Unfiltered & STD  & \multicolumn{1}{c|}{-1.407}      & 0.0000   & \multicolumn{1}{c|}{-1.222}     & 0.0000  & \multicolumn{1}{c|}{-1.3704}    & 0.0000  \\ \hline
Unfiltered & EXTRA    & \multicolumn{1}{c|}{-2.843}      & 0.0000   & \multicolumn{1}{c|}{-2.398}     & 0.0000  & \multicolumn{1}{c|}{-2.1806}    & 0.0000  \\ \hline
Unfiltered & SG  & \multicolumn{1}{c|}{-2.722}      & 0.0000   & \multicolumn{1}{c|}{-2.829}     & 0.0000  & \multicolumn{1}{c|}{-2.9815}    & 0.0000  \\ \hline
Unfiltered & FIR  & \multicolumn{1}{c|}{-4.537}      & 0.0000   & \multicolumn{1}{c|}{-4.380}     & 0.0000  & \multicolumn{1}{c|}{-4.25}      & 0.0000  \\ \hline
Unfiltered  & IIR  & \multicolumn{1}{c|}{-3.296}      & 0.0000   & \multicolumn{1}{c|}{-2.921}     & 0.0000  & \multicolumn{1}{c|}{-2.4676}    & 0.0000  \\ \hline
STD    & EXTRA    & \multicolumn{1}{c|}{-1.435}      & 0.0000   & \multicolumn{1}{c|}{-1.176}     & 0.0000  & \multicolumn{1}{c|}{-0.81019}   & 0.0001  \\ \hline
STD    & SG  & \multicolumn{1}{c|}{-1.315}      & 0.0000   & \multicolumn{1}{c|}{-1.607}     & 0.0000  & \multicolumn{1}{c|}{-1.6111}    & 0.0000  \\ \hline
STD    & FIR  & \multicolumn{1}{c|}{-3.130}      & 0.0000   & \multicolumn{1}{c|}{-3.157}     & 0.0000  & \multicolumn{1}{c|}{-2.8796}    & 0.0000  \\ \hline
STD    & IIR  & \multicolumn{1}{c|}{-1.889}      & 0.0000   & \multicolumn{1}{c|}{-1.699}     & 0.0000  & \multicolumn{1}{c|}{-1.0972}    & 0.0000  \\ \hline
EXTRA  & SG  & \multicolumn{1}{c|}{0.120 \cellcolor{yellow!85}}       & 0.9853 \cellcolor{yellow!85}   & \multicolumn{1}{c|}{-0.431\cellcolor{yellow!85}}     & 0.1590\cellcolor{yellow!85}  & \multicolumn{1}{c|}{-0.80093}   & 0.0001  \\ \hline
EXTRA  & FIR  & \multicolumn{1}{c|}{-1.694}      & 0.0000   & \multicolumn{1}{c|}{-1.982}     & 0.0000  & \multicolumn{1}{c|}{-2.0694}    & 0.0000  \\ \hline
EXTRA  & IIR  & \multicolumn{1}{c|}{-0.454\cellcolor{yellow!85}}      & 0.1181\cellcolor{yellow!85}   & \multicolumn{1}{c|}{-0.523}     & 0.0426  & \multicolumn{1}{c|}{-0.28704\cellcolor{yellow!85}} & 0.6022 \cellcolor{yellow!85}  \\ \hline
SG                  & FIR  & \multicolumn{1}{c|}{-1.815}      & 0.0000   & \multicolumn{1}{c|}{-1.551}     & 0.0000  & \multicolumn{1}{c|}{-1.2685}    & 0.0000  \\ \hline
SG                  & IIR  & \multicolumn{1}{c|}{-0.574}      & 0.0179   & \multicolumn{1}{c|}{-0.093\cellcolor{yellow!85}}     & 0.9956\cellcolor{yellow!85}  & \multicolumn{1}{c|}{0.51389}    & 0.0493  \\ \hline
FIR    & IIR  & \multicolumn{1}{c|}{1.241}       & 0.0000   & \multicolumn{1}{c|}{1.458}      & 0.0000  & \multicolumn{1}{c|}{1.7824}     & 0.0000  \\ \hline
\end{tabular}
\begin{tablenotes}
 \item[*] Highlighted values were not statistically significant.
 \item[\dag] In all case df=5, p~$<$~0.0001.
\end{tablenotes}
\end{threeparttable}
\end{table*}

\subsection{Illustration of the Effects of Filtering on Positional Signal and Velocity}
\label{exemplar}

\begin{figure*}[htbp]
\centering
\includegraphics[width=1\textwidth]{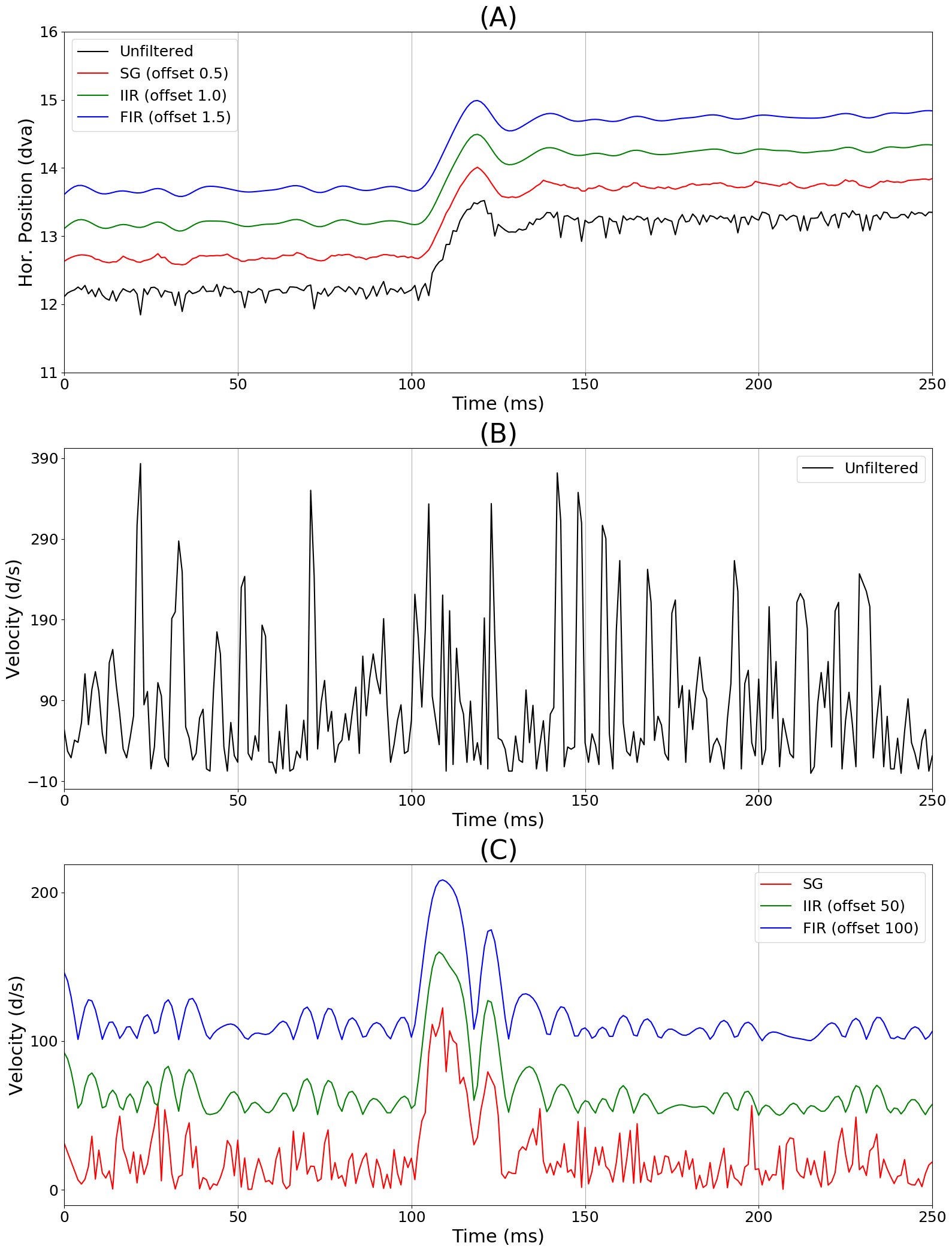}
\caption{Illustration of the effect of filtering on positional signals and instantaneous velocity.  
 A very noisy stretch of recording during our random saccade task was chosen.  (A) Horizontal position signal, including a saccade of $\approx 1.25$ degrees of visual angle (dva) for a very noisy unfiltered recording and for filtered position signals of the same recording.  Each of the filtered versions has an offset for better visualization. (B) Velocity (instantaneous) channel for the unfiltered data. (C) Velocity (instantaneous) channel of the filtered data.  The IIR velocity channel was offset by 50 degrees per second (d/s) and the FIR velocity channel was offset by 100 d/s.}
\label{fig:examplar}
\end{figure*}

In Fig.\ref{fig:examplar}, we illustrate the effect of three filtering schema (SG, IIR, FIR) on a particularly noisy unfiltered segment from our random saccade test. We do not have a method to apply the heuristic STD and EXTRA filters for this analysis since these filter functions are proprietary.   Plot (A) shows the effect of filters on the raw signal. Plot (B) represents instantaneous velocity  for the unfiltered position channel in (A). Plot (C) shows the effect of our three filters on the instantaneous velocity of the filtered signals in (A).  The IIR and FIR filters are much more effective at reducing noise than the SG filter.  The differences between the FIR and the IIR are very subtle.

%% file: sections/04-discussion.tex
\section{Discussion}
\label{discuss}

In our prior paper \citep{raju_signal} we determined that sine-waves below 100 Hz comprise signal and sine-waves above 100 Hz comprise noise.  In this paper, We compared the frequency response of 5 filters applied to eye-movement fixation signals recorded from an EyeLink 1000 eye-tracking device.  We conclude that, if the goal of the filtering process is to retain signal and remove noise, then our FIR filter is the best. This is apparent in the frequency-response and amplitude spectra of the various filtered signals.  It is also supported by a visual inspection of a particularly noisy recording with a saccade.  A large majority (82\%) of unfiltered signals exhibit statistically significant temporal autocorrelation, but all our filters substantially increase temporal autocorrelation. If the choice of a filter is based on a desire to impart the least additional temporal autocorrelation, then the heuristic STD filter is the best.

Of course, some may not agree with our 100 Hz cutoff for distinguishing between signal and noise. Although the heuristic filters have no input parameters, the SG, the FIR, and the IIR filter can be designed to have any reasonable cutoff.  

Although the FIR filter was the best, the IIR filter (low pass, \nth{7} order Butterworth) also performed very well.  The heuristic filters did tend to reduce signals above 100 Hz, but the roll-off of these filters was very gradual and shallow.  The SG filter is undesirable because of its relatively slow, shallow roll-off and because of the large ringing in the stop band.

\cite{stampe} promoted heuristic filters in place of other linear filters. He suggested that digital low-pass filters would negatively affect saccade detection but he did not provide any evidence for this claim.  It seems to us that the digital filters we propose would improve event detection because the saccade shape would be preserved and noise would be reduced.  For example, we think that event detection would be substantially easier in the filtered data in Fig. \ref{fig:examplar} than in the unfiltered data.  However, this remains an empirical question.

We studied the effect of filtering on temporal autocorrelation. As noted above, the unfiltered signals were generally temporally autocorrelated.  The lag 1 autocorrelation (ACF) was $\approx 0.58$.  The lag 1 ACF for the STD filter was $\approx 0.95$, and all the remaining filters had lag 1 ACFs $\approx 0.97$.  The marked increase in temporal autocorrelation as a result of our filters was not surprising \citep{autocorrelation} (see also footnote 4 above).  We consider the presence of temporal autocorrelation to be undesirable.  It is possible that different eye trackers might induce lower temporal autocorrelation.  This has not been studied.  Although there are time-series models (ARIMA-type models) that can markedly reduce or eliminate temporal autocorrelation, it is very unlikely (based on some pilot work) that the non-autocorrelated signals produced by such models would be useful to those who study eye movements.  So, at least for now, eye-movement researchers will have to live with the presence of temporal autocorrelation.  

We recommend that EyeLink users collect their data unfiltered at 1000 Hz.  We further recommend that EyeLink 1000 users low-pass filter their data with our FIR filter. In this way, they will retain the needed signal and remove noise.

In the future, it might be interesting to perform the same type of study using other popular eye-tracking devices. Perhaps our analysis would yield different results for different systems.  For the present, our results apply to EyeLink 1000 eye-trackers only.